\documentclass[final,author year,3p,times]{elsarticle}

\usepackage{color}
\usepackage{amsmath}
\usepackage{amssymb}
\usepackage{graphicx}
\usepackage{grffile}
\usepackage{epstopdf}
\usepackage{setspace}
\usepackage{subfigure}
\usepackage{natbib}
\usepackage{lscape}
\usepackage{booktabs}
\graphicspath{{./Figs/}}
\usepackage{upgreek}
\usepackage{nomencl}
\usepackage{multicol}
\makenomenclature

\newcommand{\Rey}{{\rm Re}}
\newcommand{\eff}{{\it eff}}

\usepackage{amssymb}

\journal{Fluids and Structures}

\begin{document}


\begin{frontmatter}



\title{Harnessing Electrical Power from Vortex-Induced Vibration of a Circular Cylinder}


\author[1]{Atul Kumar Soti}
\author[2]{Mark C. Thompson}
\author[2]{John Sheridan}
\author[3]{Rajneesh Bhardwaj}

\address[1]{IITB-Monash Research Academy, IIT Bombay, Mumbai, Maharashtra, 400076, India}
\address[2]{Fluids Laboratory for Aeronautical and Industrial Research (\textit{FLAIR}),\\
Department of Mechanical and Aerospace Engineering, Monash University, Clayton 3800, Australia}
\address[3]{Department of Mechanical Engineering, IIT Bombay, Mumbai, Maharashtra, 400076, India}

\begin{abstract}
Renewable energy sources are likely to become essential due to continuously increasing energy demands together with the depletion of natural resources that are currently used for power generation, such as coal and gas. They are also advantageous in terms of their reduced environmental impact. Here, the generation of electrical power from Vortex-Induced Vibration (VIV) of a cylinder is investigated numerically. The cylinder is free to oscillate in the direction transverse to the incoming flow. The cylinder is attached to a magnet that can move along the axis of a coil made from conducting wire. The magnet and the coil together constitute a basic electrical generator. When the cylinder undergoes VIV, the motion of the magnet creates a voltage across the coil, which is connected to a resistive load. By Lenz's law, induced current in the coil applies a retarding force to the magnet. Effectively, the electrical generator applies a damping force on the cylinder with a spatially varying damping coefficient. For the initial investigation reported here, the Reynolds number is restricted to $Re \leq 200$, so that the flow is laminar and two-dimensional (2D). The incompressible 2D Navier-Stokes equations are solved using an extensively validated spectral-element based solver. The effects of the electromagnetic (EM) damping constant $\xi_m$, coil dimensions (radius $a$, length $L$), and mass ratio on the electrical power extracted are quantified. It is found that there is an optimal value of $\xi_m$ ($\xi_{opt}$) at which maximum electrical power is generated. As the radius or length of the coil is increased, the value of $\xi_{opt}$ is observed to increase. Although the maximum average power remains the same, a larger coil radius or length results in a more robust system in the sense that a relatively large amount of power can be extracted when $\xi_m$ is far from $\xi_{opt}$, unlike the constant damping ratio case. The average power output is also a function of Reynolds number, primarily through the increased maximum oscillation amplitude that occurs with increased Reynolds number at least within the laminar range, although the general qualitative findings seem likely to carry across to high Reynolds number VIV.

\end{abstract}


\end{frontmatter}

\nomenclature{$D$}{Cylinder diameter}
\nomenclature{$L_c$}{Cylinder length}
\nomenclature{$y$}{Transverse displacement of the cylinder}
\nomenclature{$\dot{y}$}{Transverse velocity of the cylinder}
\nomenclature{$\ddot{y}$}{Transverse acceleration of the cylinder}
\nomenclature{$L$}{Length of the conducting coil}
\nomenclature{$a$}{Radius of the conducting coil}
\nomenclature{$N$}{Number of turns in the conducting coil}
\nomenclature{$y_{cm}$}{Distance between the magnet and coil}
\nomenclature{$U$}{Free stream velocity}
\nomenclature{$U_r$}{Reduced velocity}
\nomenclature{$\nu$}{Kinematic viscosity of the fluid}
\nomenclature{$\mu_m$}{Magnetic moment of the magnet}
\nomenclature{$Re$}{Reynolds number}
\nomenclature{$\mathbf{u}$}{Fluid velocity vector}
\nomenclature{$p$}{Fluid kinematic pressure}
\nomenclature{$\rho$}{Fluid density}
\nomenclature{$m_{cm}$}{Total mass of the cylinder-magnet assembly}
\nomenclature{$m$}{Mass ratio}
\nomenclature{$m_{f}$}{Mass of the displaced fluid}
\nomenclature{$C_{L}$}{Lift coefficient}
\nomenclature{$F_{m}$}{Electromagnetic force}
\nomenclature{$\xi_{m}$}{Electromagnetic damping ratio}
\nomenclature{$\xi_{m0}$}{Electromagnetic damping constant}
\nomenclature{$\xi$}{Damping ratio}
\nomenclature{$k$}{Stiffness of the spring}
\nomenclature{$c$}{Damping coefficient for the linear damping}
\nomenclature{$c_m$}{Damping coefficient for the electromagnetic damping (= $c_{m0}g^2$)}
\nomenclature{$c_{m0}$}{Electromagnetic constant ( = $\mu_m^2 /(RD^4)$)}
\nomenclature{$g$}{A non-linear function of $a$, $L$, $y_{cm}$ and $N$}
\nomenclature{$f_{n}$}{Natural frequency of the system in vacuum}
\nomenclature{$f_{N}$}{Natural frequency of the system in fluid}
\nomenclature{$P$}{Instantaneous power}
\nomenclature{$\overline{P}$}{Average power}
\nomenclature{$\overline{P}_{max}$}{Maximum average power}
\nomenclature{$R$}{Net electrical load resistance}

\printnomenclature

\section{Introduction}
Renewable energy sources, such as wind, solar, geothermal and gravitational, are receiving increased attention due to the continuing depletion of natural resources used for coal or gas-fired electrical power generation together with the associated environmental emissions.  An alternative approach to electrical power generation is to convert the available flow energy of a free-flowing fluid into electrical energy. Wind turbines provide one example \cite{carli2010high, vermeer2003wind}, although the effect of their low-frequency noise on health is debated \cite{knopper2011health}. Another example is the use of flexible piezoelectric structures. \cite{michelin2013energy} studied the efficiency of the electrical power generation due to flutter of a flexible flag with piezoelectric patches attached. They used a semi-analytical approach where the fluid forces were modelled assuming potential flow and the Euler-Bernoulli model was used for the flag. A maximum 12$\%$ efficiency was reported for the largest mass ratio and flow velocity considered in their work. \cite{akcabay2012hydroelastic} numerically studied the fully coupled fluid-solid interaction of a thin beam in incompressible viscous flow. The effect of the ratio of solid to fluid inertia forces on the flutter was investigated and the possibility of power extraction was demonstrated. \cite{akaydin2010wake} performed experiments for energy extraction from a piezoelectric beam located in the wake of a circular cylinder. The length of the beam was equal to the cylinder diameter ($D$). The power output was examined as a function of transverse and downstream beam location. Maximum power was produced when the beam was positioned on the centreline at $2D$  from the rear of the cylinder. \cite{wang2012electromagnetic} exploited pressure fluctuations in the wake of a trapezoidal bluff body to oscillate a flexible diaphragm connected to a permanent magnet. A theoretical model was developed to study the effects of the system parameters on power generation. A prototype was constructed that produced several micro-watts of electrical power.

Vortex-induced vibration (VIV) of a bluff body provides another method for extracting power from flow energy that has received considerable attention; see the review by \cite{xiao2014review} on flapping-foil based energy harvesters. \cite{mehmood2013piezoelectric} numerically investigated energy harvesting from VIV of a circular cylinder using a piezoelectric transducer. It was found that there is an optimal load resistance for harvesting maximum power but the optimal case does not coincide with the largest amplitude of oscillations. In their experiments, \cite{nishi2014power} considered two identical cylinders separated by a fixed distance in the streamwise direction. Two cases, for which one then the other cylinder was kept fixed, were compared with the isolated cylinder case. An electromagnetic transducer was used to convert the cylinder motion into electrical energy. It was found that the case with the rear cylinder fixed produced the largest vibration amplitude and the highest efficiency of 15$\%$. \cite{dai2016orientation} experimentally investigated four different installations of a cylinder on a piezoelectric beam for harnessing energy. Out of the four, three configurations had the cylinder axis aligned with the beam and for the fourth the cylinder had its axis perpendicular to the beam. The fourth configuration was reported to produce the maximum power. \cite{hobbs2012tree} experimentally investigated energy extraction from a linear array of four cylinders. Each cylinder was attached to a piezoelectric energy transducer. They studied the effect of cylinder spacing and flow speed on extracted power. Downstream cylinders were found to produce a larger amount of power than upstream cylinders. They also found that power increased only until the third cylinder for low wind speeds but continued to increase for the fourth cylinder for higher wind speeds. This suggests that the optimal number of power harvesting devices depends on the Reynolds number, at least for a certain range. \cite{barrero2012extracting} undertook a semi-analytical analysis of power harvesting from transverse VIV of a circular cylinder. The energy extraction process was modelled as a linear damper. Data from forced vibration experiments were used as input for the governing equations of motion of the cylinder. The effects of mass and damping ratios were studied, and it was found that relatively high efficiency can be achieved over a large range of reduced velocity for lower mass ratios.

\textcolor{black} {From the above review, it is clear that there has been considerable valuable research undertaken on this important problem. Given the complexity of the system and flow physics involved, it is understandable that simplifications have been made in the modelling undertaken. To-date there has been no attempt to match the form of the extracted power output to more realistic damping models. To do so is important for two reasons --- it could affect the power output if the form of its extraction is not constant versus variable damping, and indeed the variable damping could also affect the flow physics of this vortex-induced vibration system and consequently the incoming power input available for extraction. Hence,} the validity of the assumption that power output predictions based on constant damping ratio provides a good predictor for the variable damped case has not been investigated in the literature. In addition, a detailed analysis of a magnet-coil type energy harvester based on the principle of electromagnetic induction appears yet to have been reported. The aim of the present work is to address these open questions.

The layout of this paper is as follows. In section \ref{subsec:PDM}, the governing theory for the fluid-structure interaction problem is presented, followed by the specifics of the electromagnetic energy-extraction system. The numerical approach to solve this coupled system is then presented. Key results are given in section 3. In section \ref{subsec:CvsEM} the extracted power is quantified and compared for systems having constant and electromagnetically based variable damping coefficients. Following this, the details of the EM damping system are discussed. Next the influences of tuneable parameters on system behaviour are quantified, these include: the length and radius of the coil, mass ratio, and Reynolds number. Finally the influence of using dual generators is determined in section \ref{subsec:2C}, before providing conclusions. 

\section{Problem definition and methodology}\label{subsec:PDM}
\subsection{Governing equations}

\begin{figure}
\begin{center}
\includegraphics [width= 0.3 \textwidth]{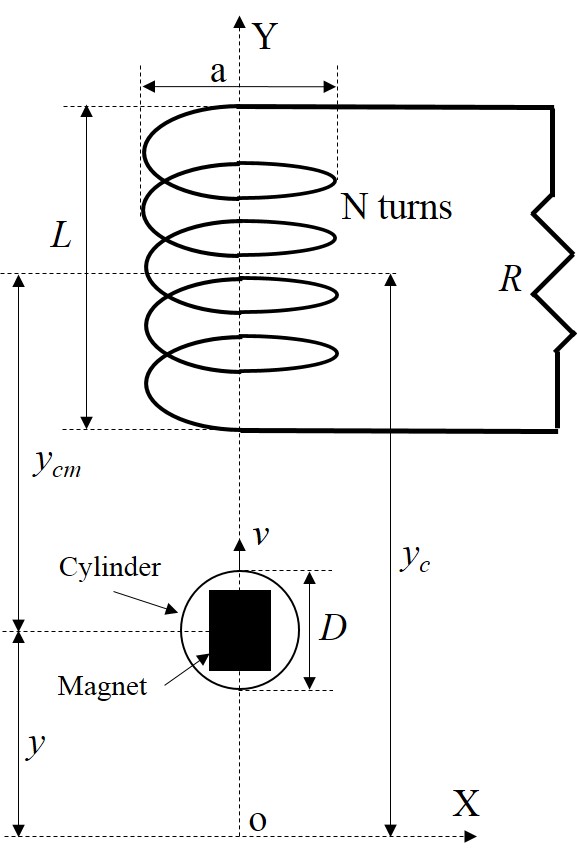}
\caption{\label{fig:1} The cylinder-magnet assembly moving along the axis of the conducting coil.}
\end{center}
\end{figure}

A vertical elastically mounted circular cylinder of diameter $D$ \textcolor{black} {(= 1)} and length $L_c$ is placed in a free-stream flow. The cylinder is free to oscillate in the transverse ($y$) direction, with the cylinder displacement denoted by $y$. The Reynolds number is chosen so that the flow is two-dimensional (2D), and the fluid is incompressible. The computational problem is set in the reference frame of the cylinder. The governing equations are the non-dimensional Navier-Stokes equations in an accelerated frame of reference
\begin{equation}
\frac{\partial	\mathbf{u}}{\partial t} +  \left(\mathbf{u}\cdot\nabla \right)\mathbf{u} = \nabla p + \frac{1}{Re}\left(\nabla ^2\mathbf{u}\right) + \mathbf{a_{F}},	\label{eq:ns}
\end{equation}
where $\mathbf{u}$ and $p$ are the fluid velocity and kinematic pressure, respectively, and  $\mathbf{a_{F}}$ is the acceleration of the reference frame. The freestream velocity $U$ and the cylinder diameter $D$ are used as the reference scales, respectively, and to define the Reynolds number, $\Rey = UD/\nu$, where $\nu$ is the kinematic viscosity of the fluid. The cylinder is attached to a magnet (magnetic moment $\mu_m$) that can move along the axis of an electrically conducting coil of non-dimensional radius $a$, length $L$, and composed of $N$ turns (see fig.~\ref{fig:1}). According to Faraday's law of electromagnetic induction, the motion of the magnet produces an emf ($\epsilon$) across the coil. If the coil is connected to a resistive load $R$ then the induced current in the coil is $i = \epsilon /R$. The induced current opposes the motion of the magnet by applying an electromagnetic force ($F_m$), given by the following expression
\begin{equation}	\label{eq:Fm}
F^*_m = c_{m0}g^2\dot{y}^*,	
\end{equation}
where * is used to represent dimensional variables. Also, $c_{m0} = \mu_m^2 /(RD^4)$ is a constant and $g = g(y(t))$ is a function of dimensions of the coil and its distance from the magnet. \textcolor{black} {It is given by the following equation based on the single magnetic dipole approximation} \cite{donoso2010magnetically}
\begin{equation}	\label{eq:g}
g = \left(\frac{2\pi Na^2}{L} \right)
\left[
\frac{1}{\left(a^2 + \left(y_{cm} - L/2\right)^2\right)^{3/2}} - 
\frac{1}{\left(a^2 + \left(y_{cm} + L/2\right)^2\right)^{3/2}}
\right],
\end{equation}
with $y_{cm}$ the non-dimensional distance between the magnet and coil. The electromagnetic force can be considered as a damping force with a non-constant damping coefficient $c_m = c_{m0}g^2$. The motion of the cylinder-magnet assembly is governed by the following equation
\begin{equation}	\label{eq:cm}
m_{cm}\ddot{y}^* + 	c\dot{y}^* + ky^* = F_l^* + F_m^*,
\end{equation}
where $m_{cm}$ is the mass of the cylinder-magnet assembly and $F_l^*$ is the lift force on the cylinder. \textcolor{black} {The second term on the left-hand side of eq.~\ref{eq:cm} represents a power transducer based on a constant damping (CD) assumption, i.e., $c = $ constant \cite{barrero2012extracting}. Notice that when the constant damping (CD) transducer is used then the electromagnetic damping (EMD) transducer is turned off and vice-versa.} The natural frequency of the above system in a vacuum is given by $f_n = \frac{1}{2\pi}\sqrt{k/m_{cm}}$. Following the standard procedure of nondimensionalization of eq.~\ref{eq:cm}, the damping and mass ratios are defined as $\xi = c/c_c$ and $m = m_{cm}/m_f$, respectively, where $m_f = \rho\frac{\pi}{4} D^2 L_c$ is the mass of the displaced fluid and $c_c = 4\pi m_{cm}f_n$ is the critical damping, below which the system response has a decaying sinusoidal variation. Similarly, the electromagnetic damping constant is defined as $\xi_{m0} = c_{m0}/c_c$, so that the non-dimensional expression for the electromagnetic force is given by
\begin{equation}	\label{eq:Fm_nd}
F_m = 2\pi^2 m \xi_m f_n \dot{y},	
\end{equation}
with $ \xi_m = \xi_{m0}g^2 $ the EM damping ratio. Using above definitions, the following non-dimensional form of eq.~\ref{eq:cm} is obtained as
\begin{equation}	\label{eq:cm_nd}
\ddot{y} + 	4\pi(\xi + \xi_m) f_n \dot{y} + 4\pi^2 f_n^2 y = \frac{2}{\pi}\frac{C_L}{m}.
\end{equation}
Here, $C_L = F_l^*/\left(\frac{1}{2}\rho U^2 D L_c\right)$ is the lift coefficient for the cylinder. In addition, the natural frequency of the cylinder-magnet assembly in the fluid is given by $f_N = \frac{1}{2\pi}\sqrt{k/m_{{\eff}}}$. The {\em effective mass} of the system in the fluid ($m_{\eff}$) is given by the sum $m_{\eff} = m_{cm} + m_a$, where $m_a$ is the added mass of the fluid which the cylinder accelerates. The added mass can be expressed as $m_a = c_a m_f$ with $c_a = 1$ from potential flow theory. The non-dimensional reduced velocity is defined as $U_r = U/(f_N D)$. \textcolor{black} {The dimensionless power is defined as the ratio of the power dissipated by the damper to the power available over the fluid region occupied by the cylinder ($\frac{1}{2}\rho U^3DL_c$). It can also be considered as the efficiency of the system.} The electrical power can be calculated by multiplying the electromagnetic force with the velocity of the magnet $P(t) = F_m \dot{y}$. The average power over a period of oscillation ($T$) of the cylinder is calculated as
\begin{equation}	\label{eq:Pa_nd}
\overline P = \frac{1}{T}\int_T P(t)\,dt.
\end{equation}

\begin{figure}
\begin{center}
\includegraphics [width= 0.5 \textwidth]{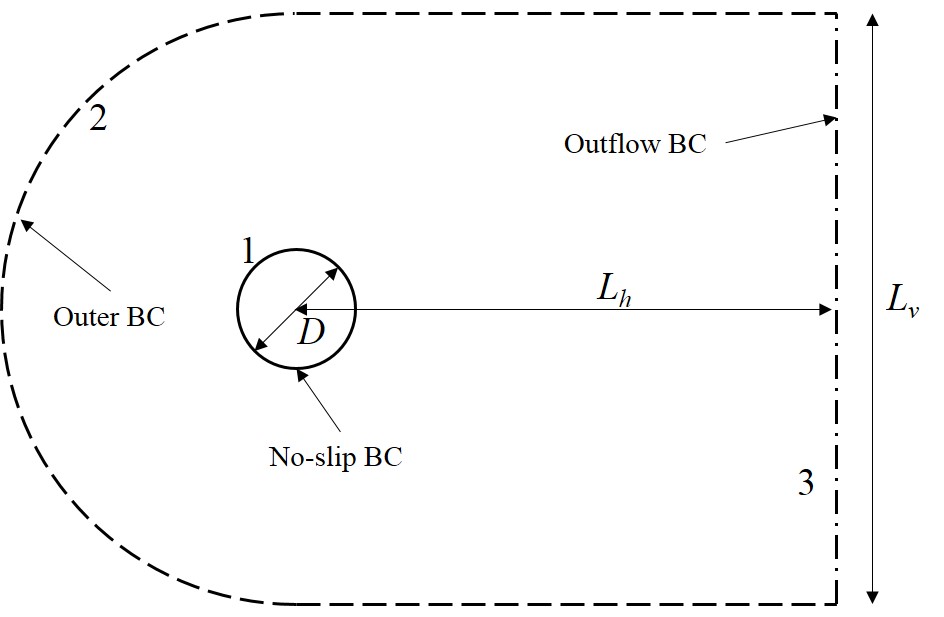}
\caption{\label{fig:2} Computational domain for VIV of a circular cylinder.}
\end{center}
\end{figure}

\subsection{Numerical formulation}
The fluid equations and cylinder equations of motion are solved in a coupled manner using a previously validated spectral-element code \cite[e.g., see ][and references therein]{LeThHo_jfs:2006, LeStThHo_pof:2006, ThHoChLe_amm:2006}. More details on the method can be found in those papers, so only a brief description will be given here. The spatial discretisation uses the nodal spectral-element approach \cite{KaSh_book:1999}, which is essentially a high-order Galerkin finite-element method. The {\it shape and weighting} functions are tensor product Lagrangian interpolating polynomials based on node points distributed according to the Gauss-Legendre-Lobatto quadrature integration points, which in turn is used to approximate the integrals from application of the weighted residual method. Importantly, the method achieves spectral (or exponential) convergence as the order of the interpolating polynomials is increased \cite{KaSh_book:1999}. The time-integration of the spatially discretised equations is achieved by a fractional step or time-splitting method \cite[e.g., ][]{Chorin_mc:1968, KaSh_book:1999, ThHoChLe_amm:2006}, in which the convective, pressure and diffusion terms of the Navier-Stokes equations are integrated sequentially, using an explicit Adam-Bashforth method for the convective substep and the $\theta$-corrected implicit Crank-Nicholson method for the diffusive substep \cite{CanHusQuaZan_book:2006}. The pressure substep is used to satisfy continuity and is also treated implicitly. The higher-order pressure boundary condition applied at the cylinder surface is derived from the Navier-Stokes equations to provide the pressure-normal derivative \cite[see][]{KaIsOr_jcp:1991}. A fuller description can be found in \cite{ThHoChLe_amm:2006}. For the coupled problem, the acceleration of the frame is added to the convective substep, and iteration proceeds during each full timestep until the fluid velocity field, cylinder velocity and the cylinder applied force converge. Testing was performed to ensure that each of these three convergence criteria are small enough to achieve 1\% accuracy or better in predicting the oscillation amplitude evolution. 
It has been previously been used to model vortex-induced vibrations of  inline \cite{LeLoTh_jfm:2013, LeLoTh_jfm:2011}, transverse (\cite{LeStThHo_pof:2006, LeThHo_jfs:2006}), and rotationally  (\cite{LoLeThSh_jfm:2010}) oscillating cylinders, and even tethered spheres \cite{LeHoTh_jfm:2013}.

\subsection{Computational domain and boundary conditions}

The computational domain, shown in fig.~\ref{fig:2}, extends $L_h = 25D$ and $L_v = 40D$ in the downstream and transverse directions, respectively. Thus the cross-stream blockage is 2.5\%. The inlet is semicircular with diameter $L_v$. The fluid velocity is prescribed as $u = U, v = -\dot{y}$ at the inlet, top and bottom boundaries, where $\dot{y}$ is the velocity of the cylinder. No-slip conditions are applied at the cylinder boundary. Neumann conditions is applied at the outlet for the fluid velocity and the pressure is taken as constant.

\subsection{Resolution studies and validation}

The base computational mesh is the same as that used by \cite{LeThHo_jfs:2006} for their studies of transverse oscillations of a circular cylinder at $\Rey = 200$, except that the blockage has been reduced by adding an extra layer of cells to extend the transverse dimension from 30 to $40D$. It was verified that the same amplitude/reduced-velocity response curve was reproduced from that study for the constant damping case. For the current paper, the majority of simulations were undertaken at $\Rey = 150$, ensuring that the flow remained two-dimensional. \cite{LeThHo_jfm:2007} showed through stability analysis that the wake does not undergo transition to three-dimensionality prior to $\Rey=250$, at least in the high-amplitude lock-in region. For the bulk of the simulations reported in this paper, $5 \times 5$ noded elements were used for the macro-elements of the mesh. This is sufficient to guarantee the prediction of the peak oscillation amplitude of vibrations to better than 1\% for the Reynolds numbers considered.

\begin{figure}
\centering   
	 \subfigure{\includegraphics[width= 0.4 \textwidth] {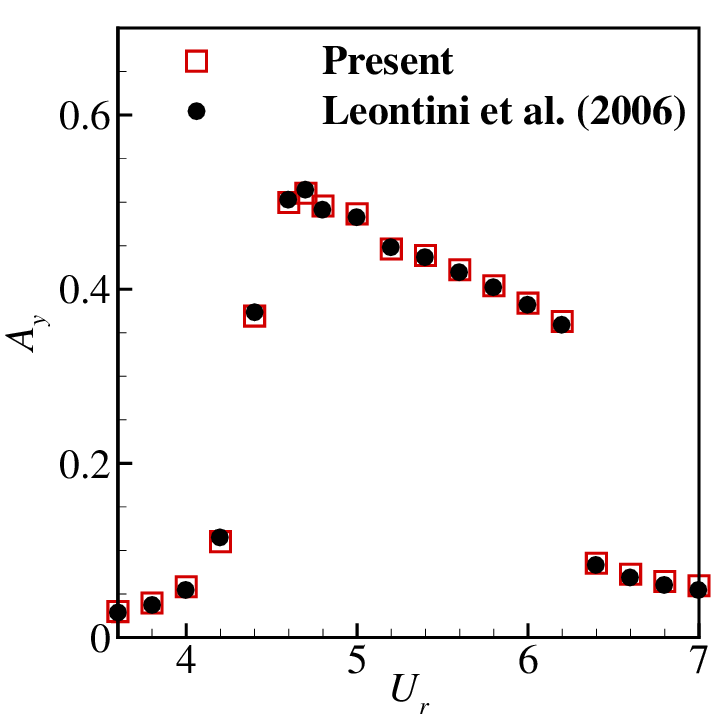}}
	 \subfigure{\includegraphics[width= 0.4 \textwidth] {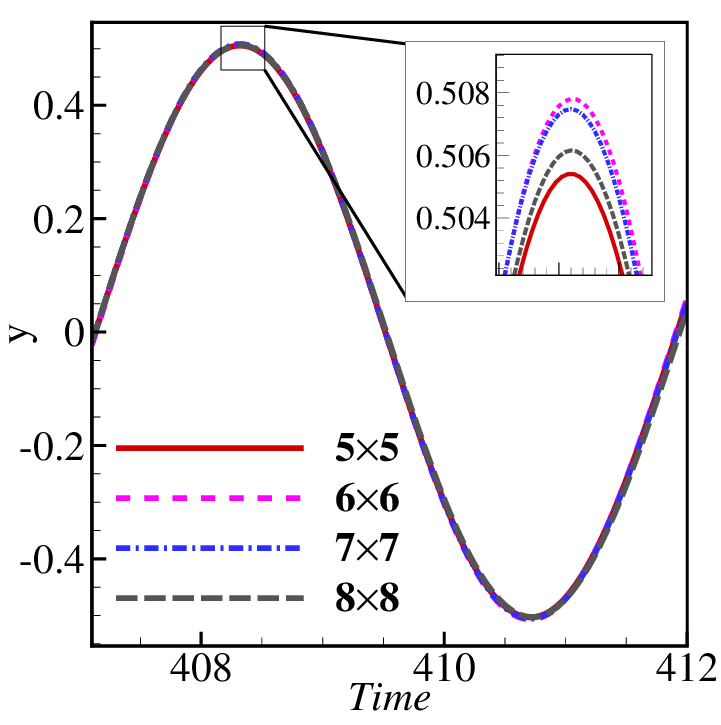}}
\caption{\label{fig:validate} a) Comparison of the cylinder response with the published data and b) effect of spectral element resolution on the cylinder displacement.}
\end{figure}

\textcolor{black} {To provide more confidence in the predictions, validation and resolution tests are presented in fig.~\ref{fig:validate}a and \ref{fig:validate}b, respectively. In fig.~\ref{fig:validate}a, the cylinder displacement amplitude ($A_y$) obtained by the present solver is compared with that of the \cite{LeThHo_jfs:2006} for a range of reduced velocities ($U_r$) at Re = 200, $m = 10$ and $\xi = 0.01$. Individual values of $A_y$ differ by less than 1\% from the published data. For the resolution study $U_r = 4.7$ was chosen, which was the reduced velocity leading to maximum cylinder displacement. To perform the resolution study, the number of elements in the computational mesh was fixed while varying the number of nodes per element. In fig.~\ref{fig:validate}b, the temporal variation of the cylinder displacement is shown for different numbers of nodes per element. This indicates the $5 \times 5$ element-based mesh predicts the maximum displacement to within better than 1\% of the most resolved mesh tested. Hence, this mesh was used for subsequent simulations. }

\section{Results}
\subsection{Constant versus EM damping ratio} \label{subsec:CvsEM}

\begin{figure}
\centering
\includegraphics[width= 0.7 \textwidth] {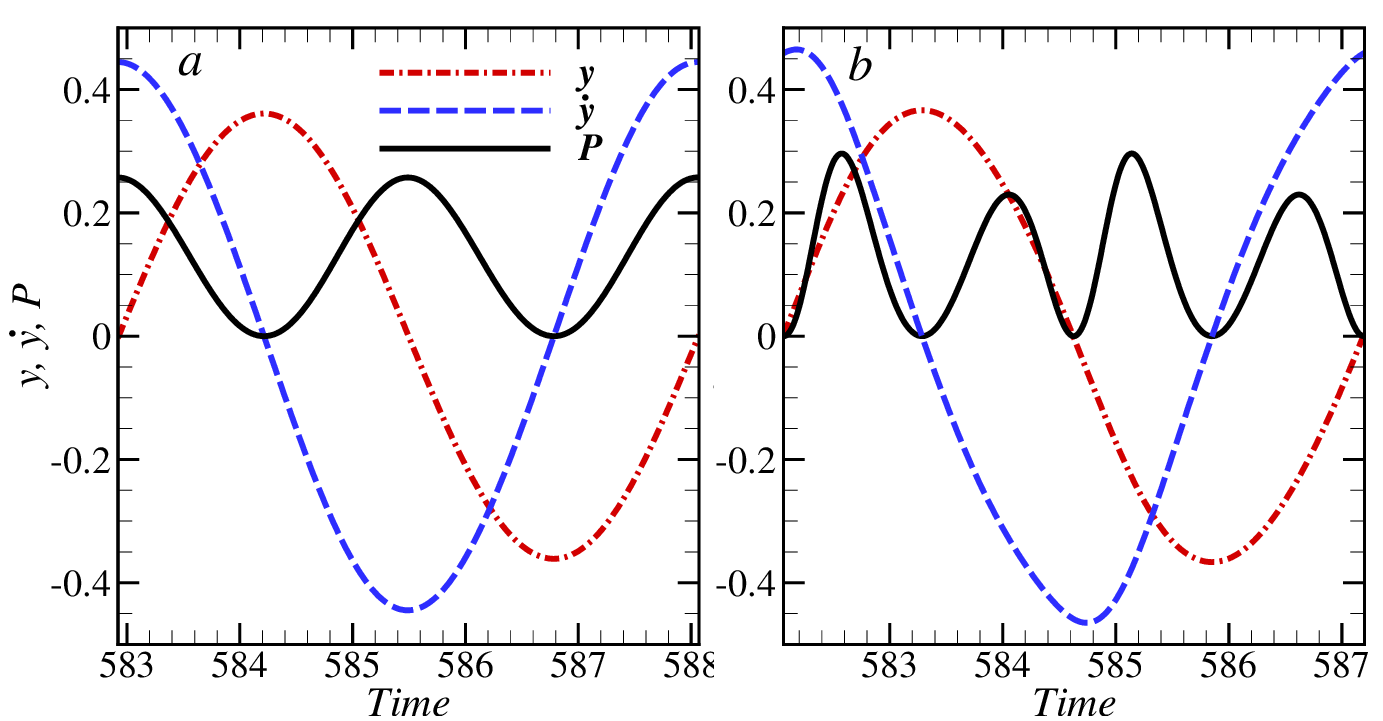}
\subfigure[]{\includegraphics[width= 0.3 \textwidth] {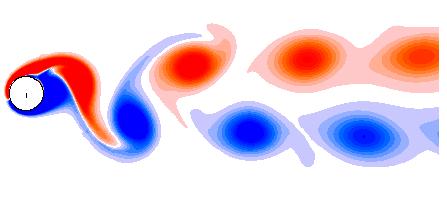}}\qquad
\subfigure[]{\includegraphics[width= 0.3 \textwidth] {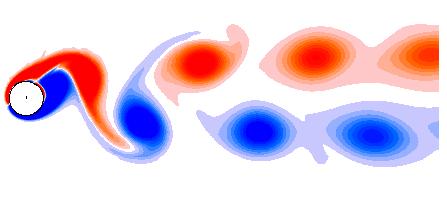}}
\caption{\label{fig:pt} Top: Temporal variation of power, position and velocity of the cylinder for (a) CD ($\xi$ = 0.14) and (b) EMD ($\xi_{m0} = 2.4 \times 10^{-5}$, $a$ = 0.6, $L$ = 0.6) cases, at $U_r$ = 5.2 and $m = 2$. Vorticity contours (scale -2 to 2) corresponding to these cases are shown below each plot.}
\end{figure}

In this section, the behaviour of using the electromagnetic damping (EMD) is compared with the constant damping (CD) in terms of energy extraction. The average power ($\overline P$) is a function of damping ratio, reduced velocity and mass ratio. Initially, the mass ratio is fixed at $m = 2$, and the reduced velocity is taken as $U_r = 5.2$ which corresponds to the lock-in condition at $Re =$ 150 \cite{LeThHo_jfs:2006}. For the EMD case, the length and radius of the coil are both taken as $0.6D$. Time variations of power, transverse displacement ($y$) and transverse velocity ($v$) of the cylinder over an oscillation cycle are shown in fig.~\ref{fig:pt} for the CD and EMD cases. The plots correspond to a value of the damping ratio that produces the maximum average power ($\overline P_{max}$). Let $f_p$ and $f_c$ be the fundamental frequencies of power and cylinder transverse displacement, respectively. As seen in fig.~\ref{fig:pt}, $f_p$ is 2 and 4 times $f_c$ for the CD and EMD cases, respectively. In the CD case, peak electrical power is generated when the cylinder passes through its mean position ($y = 0$), where it has the maximum speed. The same is not true for the EMD case, where peak power is generated when the cylinder displacement is $y = \pm 0.23 $. There are two distinct local maxima for power, as seen in fig.~\ref{fig:pt}b. Also notice that the velocity profile of the cylinder is not sinusoidal in the EMD case. The instant of maximum velocity does not coincide with that of the mean position. Therefore the two peaks of power have different magnitude even though they occur when the cylinder is at the same distance from its mean position. The lower and higher peaks occur when the cylinder is moving towards and away from its mean position, respectively. The peak power in the EMD case (0.30) is higher than that of the CD case (0.26).

The vorticity patterns for both the cases are also compared in fig.~\ref{fig:pt}. In both cases the $2S$ vortex shedding mode is observed \cite[e.g.,][]{williamson2004vortex}, and indeed there is little difference between the shedding patterns.

\begin{figure}
\centering   
	 \subfigure{\includegraphics[width= 0.4 \textwidth] {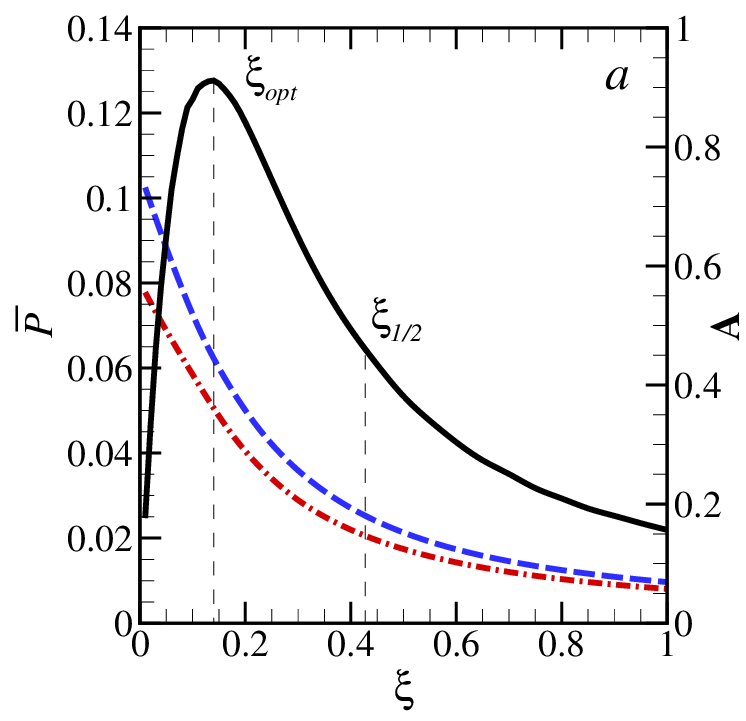}}
	 \subfigure{\includegraphics[width= 0.4 \textwidth] {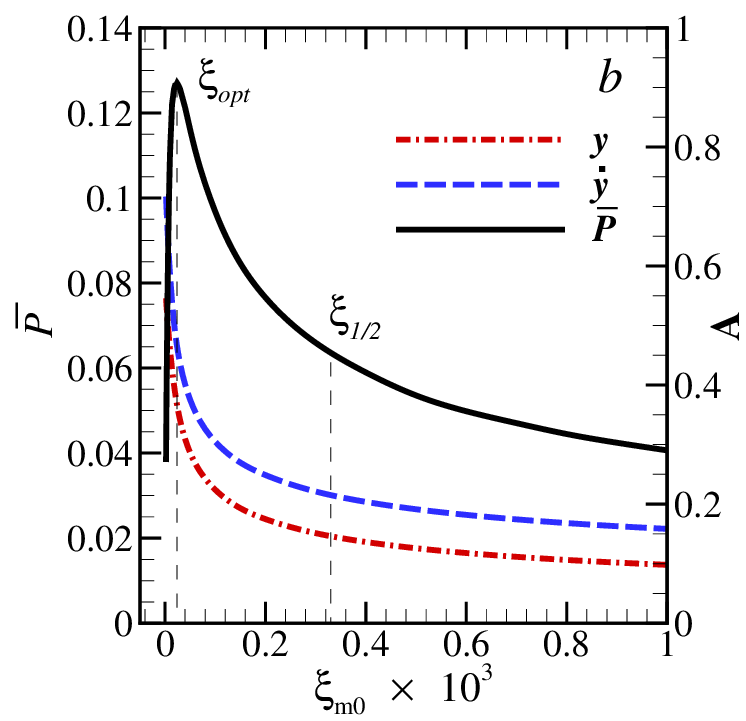}}
\caption{\label{fig:p_vs_xi} Variation of the average power (a) with $\xi$ for CD and (b) with $\xi_{m0}$ for EMD ($a = 0.6, L = 0.6$) cases, for $U_r = 5.2$ and $m = 2$.}
\end{figure}

Figure \ref{fig:p_vs_xi}a shows the variation of the average power with damping ratio ($\xi$) for the CD case. Also plotted are the displacement and velocity amplitudes of the cylinder, which are seen to  monotonically decrease with increasing $\xi$. This behaviour is expected since the role of damping is to apply a retarding force on the cylinder, hence, a larger damping value results in a larger retarding force and thereby a smaller oscillation amplitude. Since the oscillation amplitude of the cylinder decreases with $\xi$, and the power is proportional to the product of $\xi$ and the square of the velocity amplitude, it is expected that there is an optimal value of $\xi$ ($\xi_{opt}$) at which maximum power is harnessed. This is seen in fig.~\ref{fig:p_vs_xi}a, where $\xi_{opt} = 0.14$ and the maximum average power ($\overline P_{max}$) is 0.13. A similar plot is shown in fig.~\ref{fig:p_vs_xi}b for the EMD case. The maximum average power in this case is also close to 0.13, but note that more power is obtained compared to the CD case when the damping ratio is greater than its optimal value. To quantify this effect, the {\em quality} of the system is defined as
\begin{equation}	\label{eq:Q}
Q = \frac{\Delta \xi}{\xi_{opt}}.	
\end{equation}
Here $\Delta \xi$ is the half-width at half-maximum (HWHM), \textcolor{black} {i.e., $\Delta \xi = \xi_{1/2} - \xi_{opt}$, where $\xi_{1/2}$ represents the value of damping at which the power is half of its maximum value.} In effect, $Q$ signifies how far, relative to the optimal condition, the system can operate and still produce more than half of the maximum average power. The values of $Q$ at $U_r = 5.2$ for CD and EMD cases are 2.1 and 12.8, respectively. \textcolor{black} {The value of maximum average power at $U_r = 6.7$ is close to half of $\overline P_{max}$ at $U_r = 5.2$. The values of $Q$ at $U_r = 6.7$ for the CD and EMD cases are 4.1 and 31.0, respectively.} Thus, the EMD system is considerably less sensitive to tuning.

\begin{figure}
\centering
\includegraphics[width= 0.7 \textwidth] {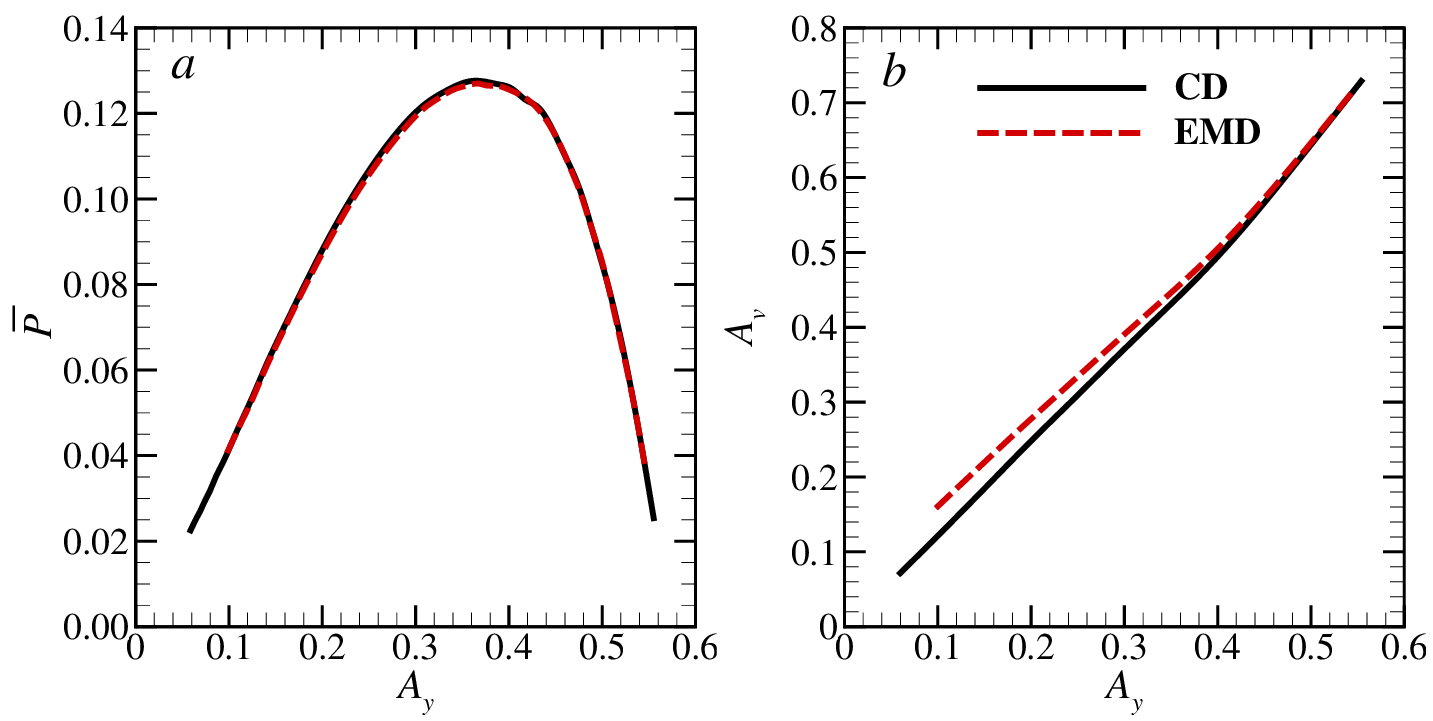}
\caption{\label{fig:p_vs_A} (a) Average power and (b) velocity amplitude versus displacement amplitude of the cylinder for CD and EMD cases at $U_r$ = 5.2, $m$ = 2 and $Re =$ 150.}
\end{figure}

In fig.~\ref{fig:p_vs_A}, the data of fig.~\ref{fig:p_vs_xi} is replotted with different axes. This reveals the relationship of power and velocity amplitude with the displacement amplitude of the cylinder for both the cases. While fig.~\ref{fig:p_vs_A}a shows the similarity between the two cases, fig.~\ref{fig:p_vs_A}b points out the differences. Although the temporal variation of power is different for each case, surprisingly, the average power is seen to depend only on the displacement amplitude of the cylinder in fig.~\ref{fig:p_vs_A}a. There is an optimal amplitude for the maximum average power. On the other hand, the velocity amplitude of the cylinder is larger for the EMD case at smaller displacement amplitude. This happens because the EMD case has smaller damping at the mean position ($y = 0$), which allows a larger acceleration of the cylinder.

\begin{figure}
\centering
\includegraphics[width= 0.4 \textwidth] {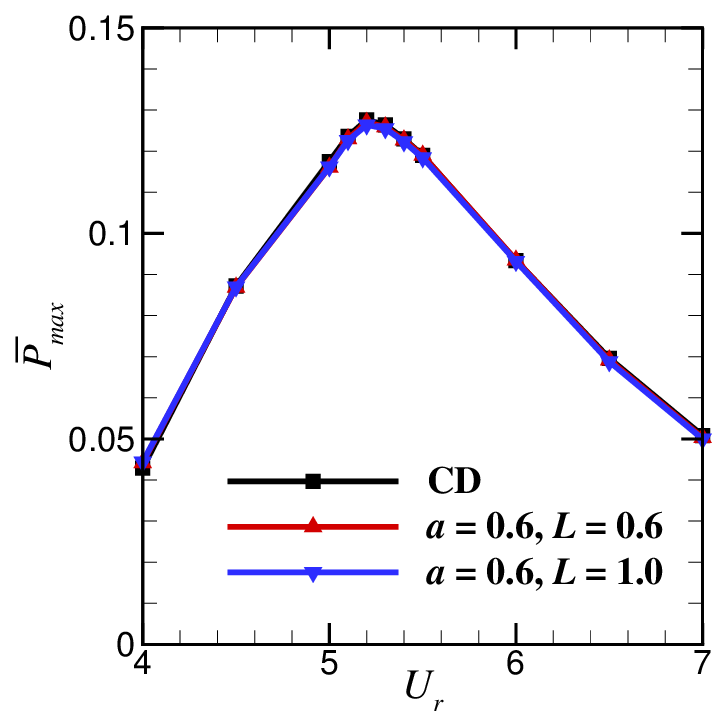}
\caption{\label{fig:pm_m2} Maximum average power versus $U_r$ for CD and EMD cases at $m$ = 2 and $Re = 150$.}
\end{figure}

Figure \ref{fig:pm_m2} shows the variation of maximum average power ($\overline P_{max}$) with reduced velocity for both CD and EMD cases. Two EMD cases with different coil lengths are considered. All the curves in fig.~\ref{fig:pm_m2} are effectively indistinguishable implying that the variation of maximum average power with reduced velocity is independent of the nature of damping used. Therefore, a constant damping ratio can be used to calculate the {\em average} power that can be extracted from the system under non-constant electromagnetic damping; whilst noting that the temporal variation for the two cases is quite different. 

\subsection{Effect of coil length}	\label{subsec:ECL}
\begin{figure}
\centering
\includegraphics[width= 0.7\textwidth] {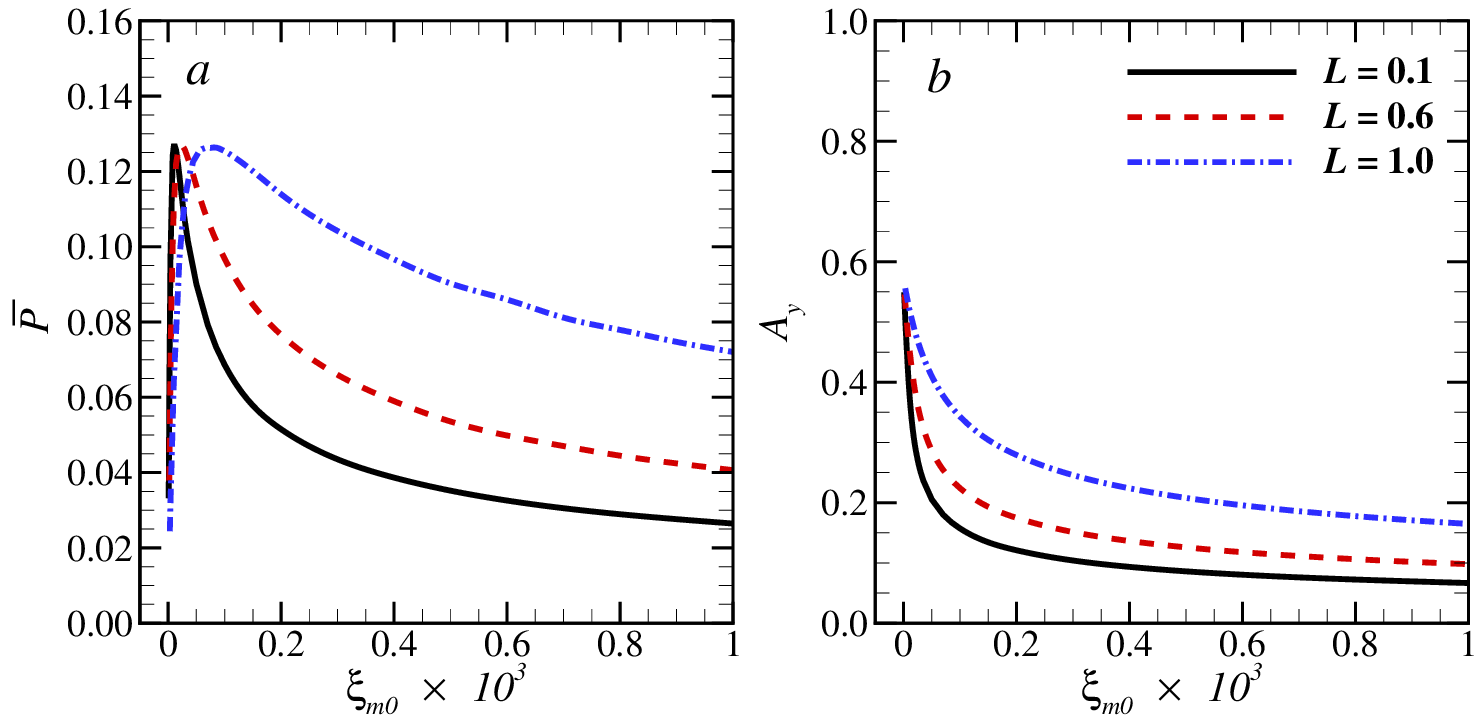}
\caption{\label{fig:p_vs_L} Effect of coil length on (a) average power and (b) displacement amplitude at $U_r$ = 5.2, $m$ = 2 and $a = 0.6$.}
\end{figure}

\begin{figure}
\centering
\includegraphics[width= 0.7\textwidth]  {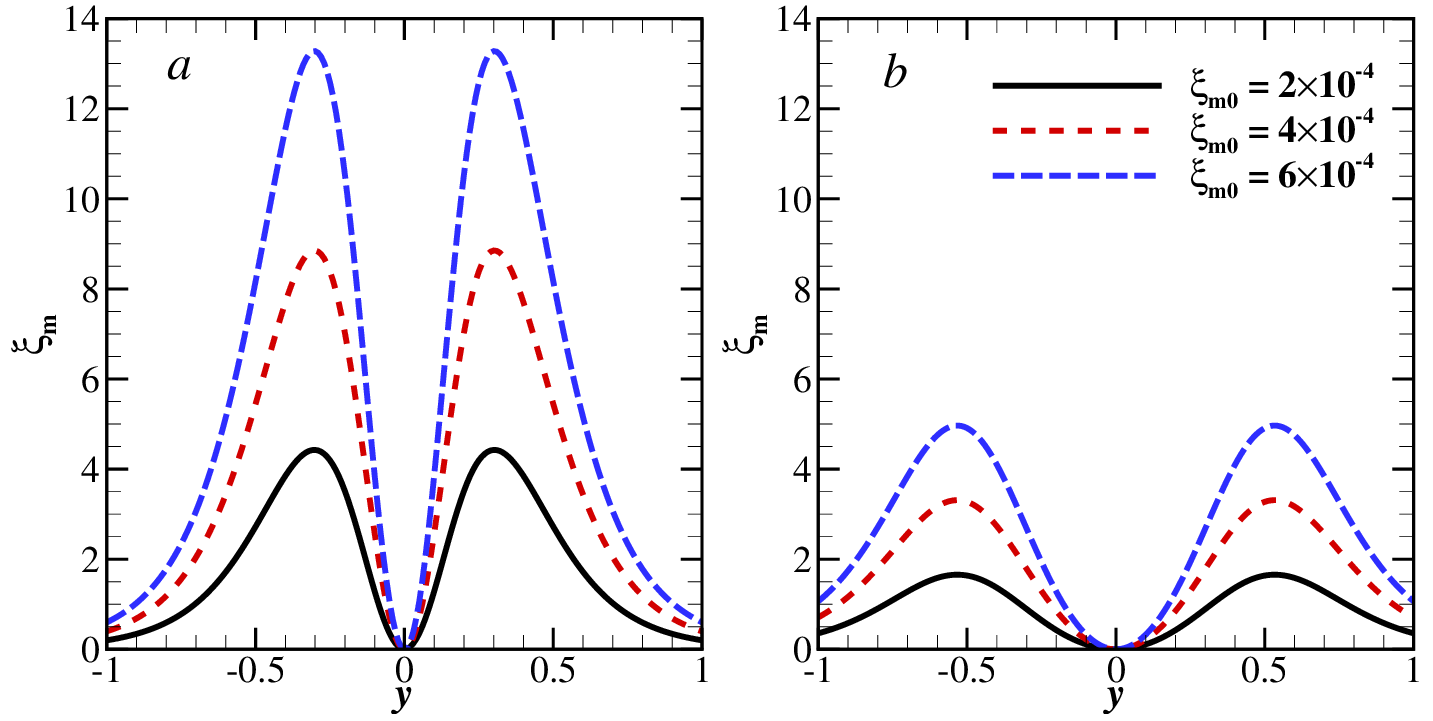}
\caption{\label{fig:xi_vs_l} Effect of coil length on EM damping ratio (a) $a = 0.6, L = 0.1$ and (b) $a = 0.6, L = 1.0$.}
\end{figure}

Now details of the EMD setup are examined while retaining the same mass ratio ($m = 2$) and choosing the lock-in condition ($U_r = 5.2$) at $Re = 150$. In this section, the effect of the length of the coil on the power output is considered. In addition to the value of 0.6 already considered, $L$ is varied in the range $[0.1, 1.0]$. The radius of the coil is kept at $a = 0.6$. The variations of average power and displacement amplitude of the cylinder with $\xi_{m0}$ are shown in fig.~\ref{fig:p_vs_L} for the different lengths of the coil ($L$). There is no significant effect of the coil length on the maximum average power, but the value of optimal $\xi_{m0}$ increases with $L$. It is worth mentioning that the relationships of power and the velocity amplitude with the displacement amplitude of the cylinder (not shown here) are unaffected by the $L$. The displacement amplitudes of the cylinder at optimal $\xi_{m0}$ for $L$ = 0.1, 0.6 and 1.0 are 0.39, 0.37 and 0.38, respectively. The corresponding velocity amplitudes of the cylinder are 0.49, 0.46 and 0.48, respectively. It is seen that at a particular value of $\xi_{m0}$, greater than the optimal value, larger values of power and velocity amplitude of the cylinder are produced for a larger $L$. This behaviour can be explained by the relationship of $\xi_{m}$ with $L$, which is plotted in fig.~\ref{fig:xi_vs_l}. In fig.~\ref{fig:xi_vs_l}a and b, the variation of $\xi_{m}$ with $y$ is plotted, at three values of $\xi_{m0}$, for $L$ = 0.1 and 1.0, respectively. As seen in fig.~\ref{fig:xi_vs_l}, a small value of $L$ results in a large value of $\xi_{m}$ at the same $\xi_{m0}$. Since increased damping ratio is expected to reduce the vibration amplitude, a small $L$ produces a smaller displacement amplitude and power at a particular value of $\xi_{m0}$. The values of $Q$ for $L$ = 0.1, 0.6 and 1.0 are 11, 12.8 and 19.6, respectively.

\subsection{Effect of coil radius}	\label{subsec:ECR}
\begin{figure}
\begin{center}
\includegraphics[width= 0.7 \textwidth] {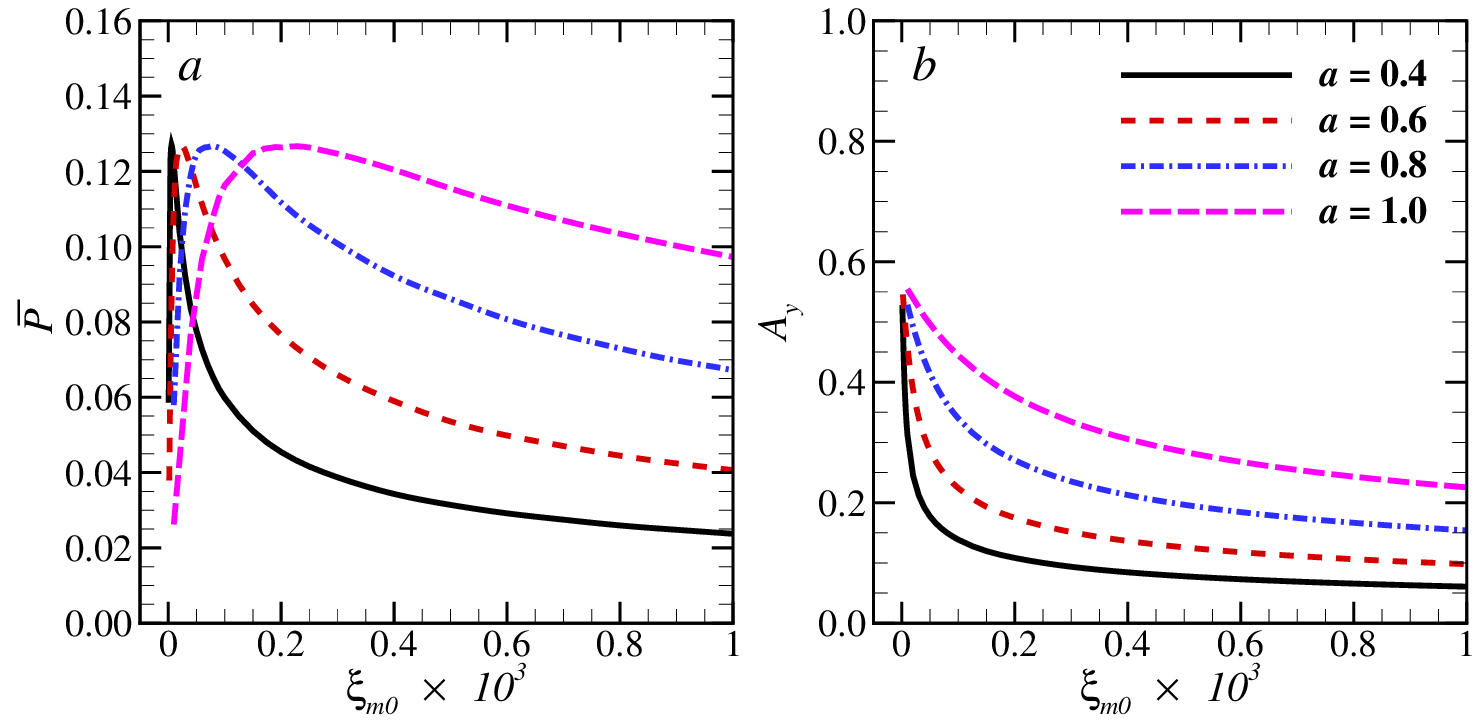}
\caption{\label{fig:p_vs_a} Effect of coil radius on (a) average power and (b) displacement amplitude at $U_r$ = 5.2, $m$ = 2 and $L = 0.6$.}
\end{center}
\end{figure}

\begin{figure}
\centering
\includegraphics[width= 0.7 \textwidth] {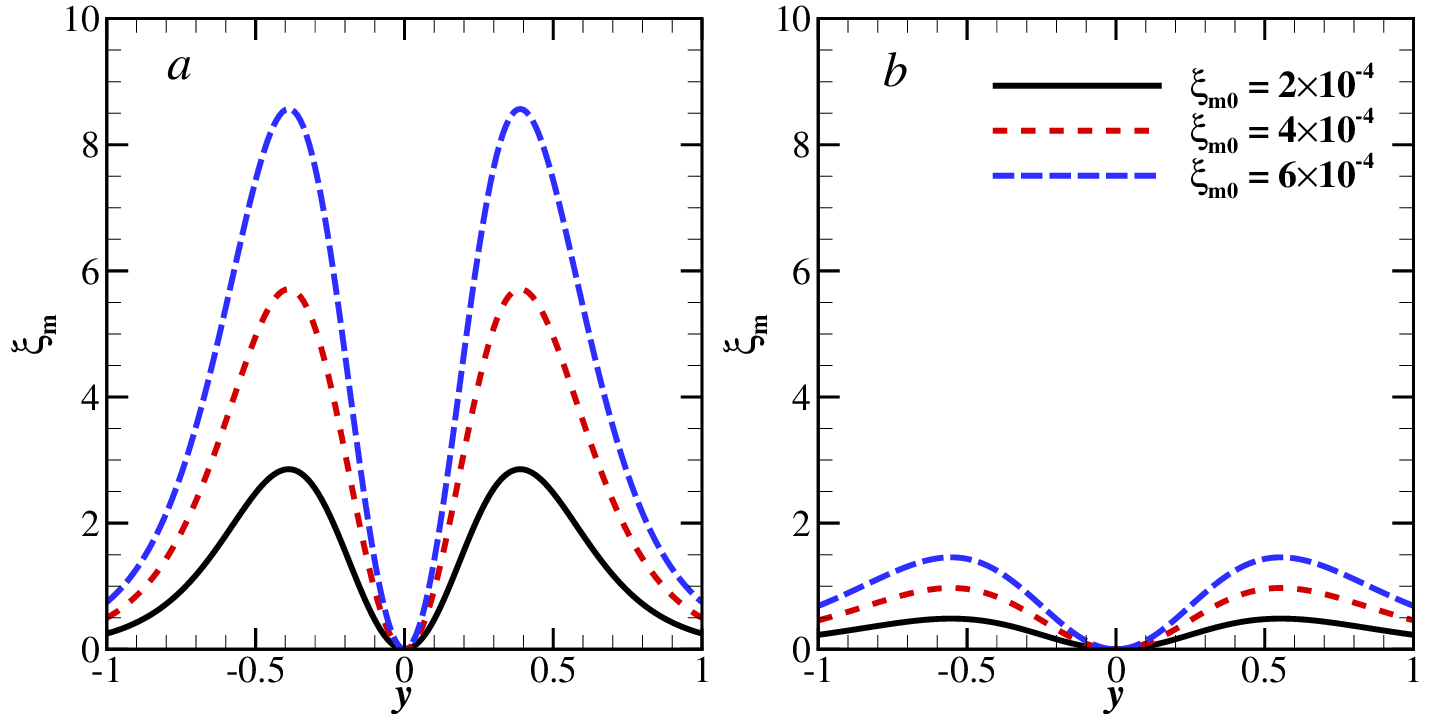}
\caption{\label{fig:xi_vs_a} Effect of coil radius on EM damping ratio (a) $a = 0.6, L = 0.6$ and (b) $a = 1.0, L = 0.6$.}
\end{figure}

This section studies the effect of the coil radius on the power output. Coil radius values of $a = 0.4$, 0.8 and 1.0 are used, in addition to the already considered value of 0.6, while the length of the coil is kept at $L = 0.6$. The other parameters are kept unchanged, i.e. $m = 2$, $U_r = 5.2$ and $Re = 150$. The effect of coil radius on power can be seen in fig.~\ref{fig:p_vs_a}, where the variations of average power and displacement amplitude of the cylinder with $\xi_{m0}$ are plotted for the four aforementioned values of $a$. Similar to what was seen in section \ref{subsec:ECL}, the value of maximum average power is unaffected by the coil radius, but the value of optimal $\xi_{m0}$ increases with an increase in $a$. The displacement amplitudes of the cylinder at optimal $\xi_{m0}$ for $a$ = 0.4, 0.6, 0.8 and 1.0 are 0.37, 0.39, 0.36 and 0.36, respectively. The effect of coil radius on damping $\xi_m$ is shown in fig.~\ref{fig:xi_vs_a} where the variation of $\xi_{m}$ with $y$ is plotted, at three values of $\xi_{m0}$, for $a$ = 0.6 and 1.0, respectively. The damping $\xi_{m}$ is higher for smaller $a$ and therefore more power is produced for larger $a$ at same $\xi_{m0}$ (fig.~\ref{fig:p_vs_a}a). The values of $Q$ for $a$ = 0.4, 0.6, 0.8 and 1.0 are 13.2, 12.8, 13.4 and 13.8, respectively. Again, the relationships of power and the velocity amplitude with the displacement amplitude of the cylinder (not shown here) are unaffected by the coil radius.

%

\subsection{Effect of mass ratio}	\label{subsec:EMR}
\begin{figure}
\begin{center}
\includegraphics[width= 0.45 \textwidth] {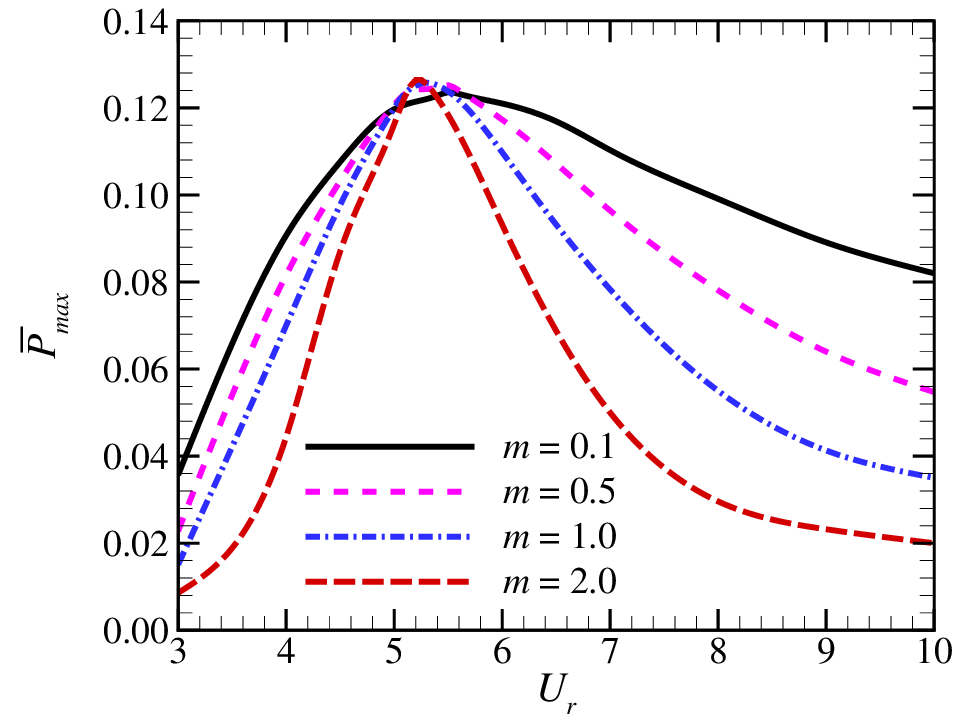}
\caption{\label{fig:pmvsm} Maximum average power versus $U_r$ for different mass ratios.}
\end{center}
\end{figure}

Next the effect of mass ratio of the cylinder-magnet assembly on the power output is discussed. In this case the coil radius ($a$) and length ($L$) are set to 0.6 and 1.0, respectively. As discussed in section \ref{subsec:CvsEM}, the values of $a$ and $L$ do not affect the maximum average power $\overline P_{max}$. Figure \ref{fig:pmvsm} shows the variation of $\overline P_{max}$ with $U_r$ for different mass ratios $m$. It is seen that the $\overline P_{max}$ versus $U_r$ curve becomes flatter as the mass ratio is decreased, implying that the synchronization region for the VIV of the cylinder becomes larger for smaller $m$. This phenomenon has been reported previously in the literature \cite{govardhan2000modes}. The results show only a very small effect of $m$ on the peak value of $\overline P_{max}$. 

\subsection{Effect of the Reynolds number}	\label{subsec:effect_Re}
\begin{figure}
\centering
\includegraphics[width= 0.7\textwidth] {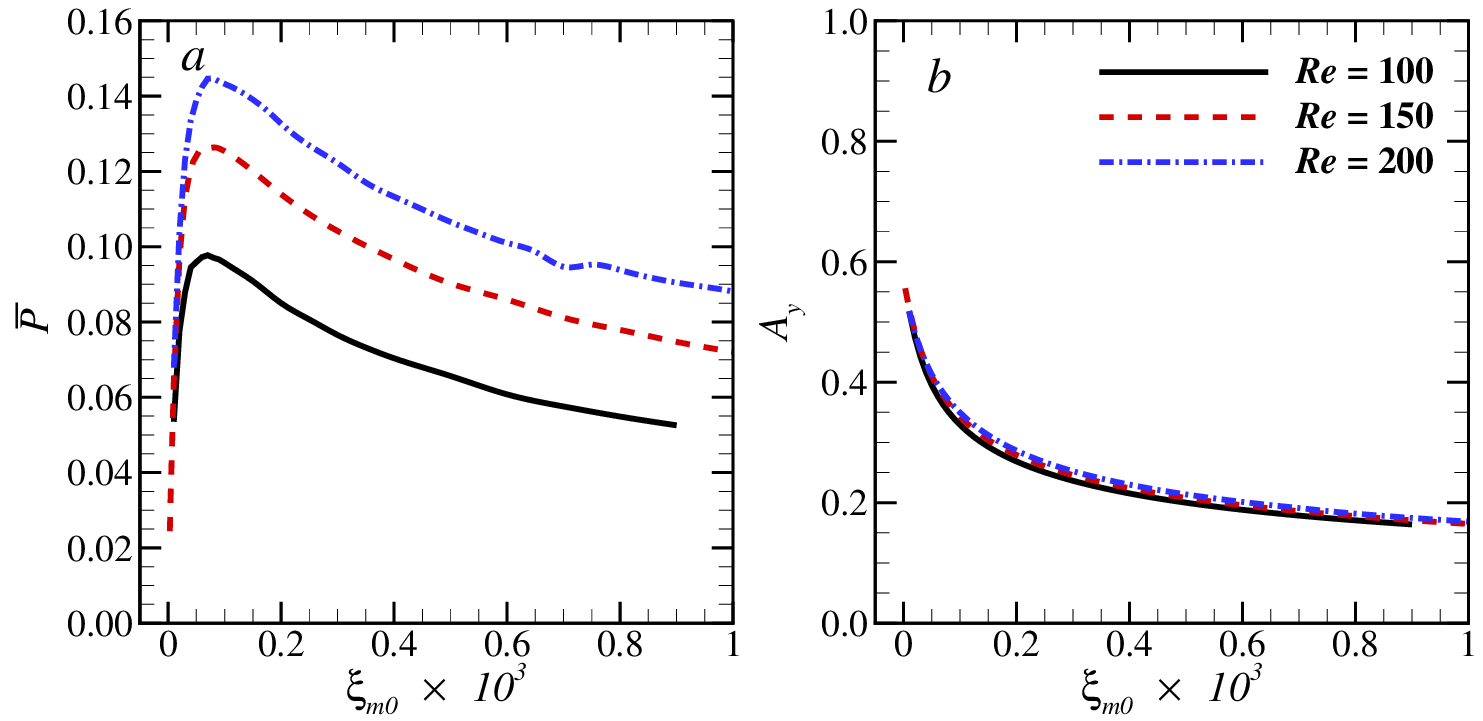}
\caption{\label{fig:p_vs_Re} Effect of Re on average power and displacement amplitude for $m$ = 2, $a = 0.6$ and $L = 1.0$.}
\end{figure}

Lastly, the effect of the Reynolds number on the power output of the system is briefly considered. For this the value of $Re$ is varied to take values 100 and 200, in addition to the value of 150 already considered. Similarly to before, the mass ratio is kept as $m = 2$, and a coil of length $L = 1.0$ and radius $a = 0.6$ is kept at $y = 0$. The variations of average power with $\xi_{m0}$ for three Reynolds numbers are shown in fig.~\ref{fig:p_vs_Re}a. The maximum average power is obtained at $U_r$ = 5.3 and 5.2 for $Re$ = 100 and 200, respectively. The average power is seen to increase with $Re$. Two reasons could account for this increase in $\overline P$ with $Re$: increase in Strouhal number and increase in the vibration amplitude (see fig.~\ref{fig:p_vs_Re}b) with $Re$. The values of 
$\overline P_{max}$ for Re = 100, 150 and 200 are 0.10, 0.13 and 0.145, respectively.

\subsection{Using two coils}	\label{subsec:2C}
In the previous sections the electrical generator consisted of only one coil, which was located at the magnet centre $y = 0$. Another possibility is to use two identical coils that are kept at equal distances, $\pm y_c$, along the transverse direction. In this case the net EM damping force is the sum of the damping forces due to each coil. Therefore, the effective EM damping ratio for the two coils case can be written as
\begin{equation} \label{eq:xi_2c}
\xi_{m} = \xi_{m0}\left(g_1^2 + g_2^2 \right).
\end{equation}
\begin{figure}
\centering
\includegraphics[width= 0.4 \textwidth] {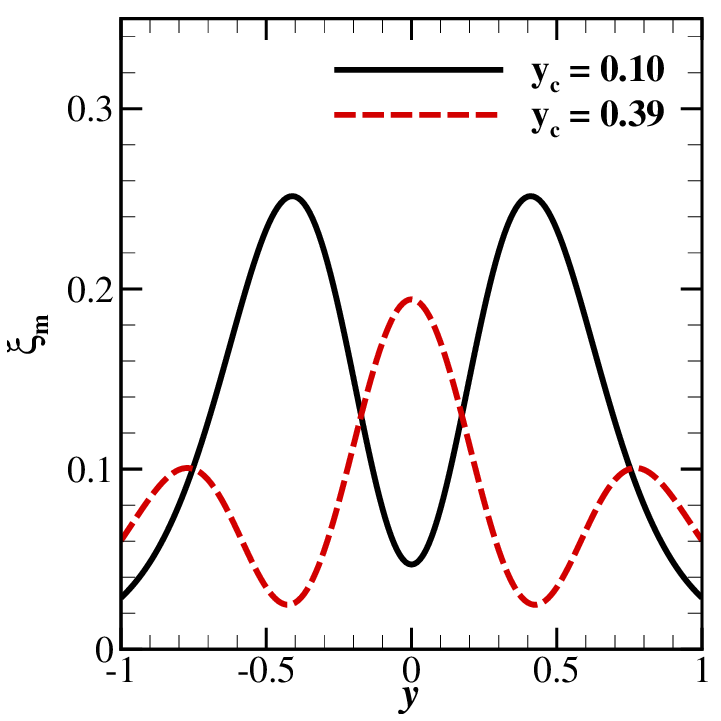}
\caption{\label{fig:xi_vs_y_2c} Variation of $\xi_{m}$ with $y$ for the two coil case. They are plotted for optimal cases ($\xi_{m0} = 1 \times 10^{-5}$ and $6.8 \times 10^{-6}$ for $y_c$ = 0.10 and 0.39, respectively).}
\end{figure}
Two values of $y_c$, 0.10 and 0.39, are considered. The variations of $\xi_{m}$ with $y$ for these two values of $y_c$ are shown in fig.~\ref{fig:xi_vs_y_2c}. The plot of $\xi_{m}$ for $y_c = 0.10$ is similar to that for one coil case, except that it has a non-zero value at the centre $y = 0$. On the other hand, the $y_c = 0.39$ case shows an entirely different behaviour where the maximum damping occurs at the centre.

\begin{figure}
\centering
\includegraphics [width= 0.7\textwidth] {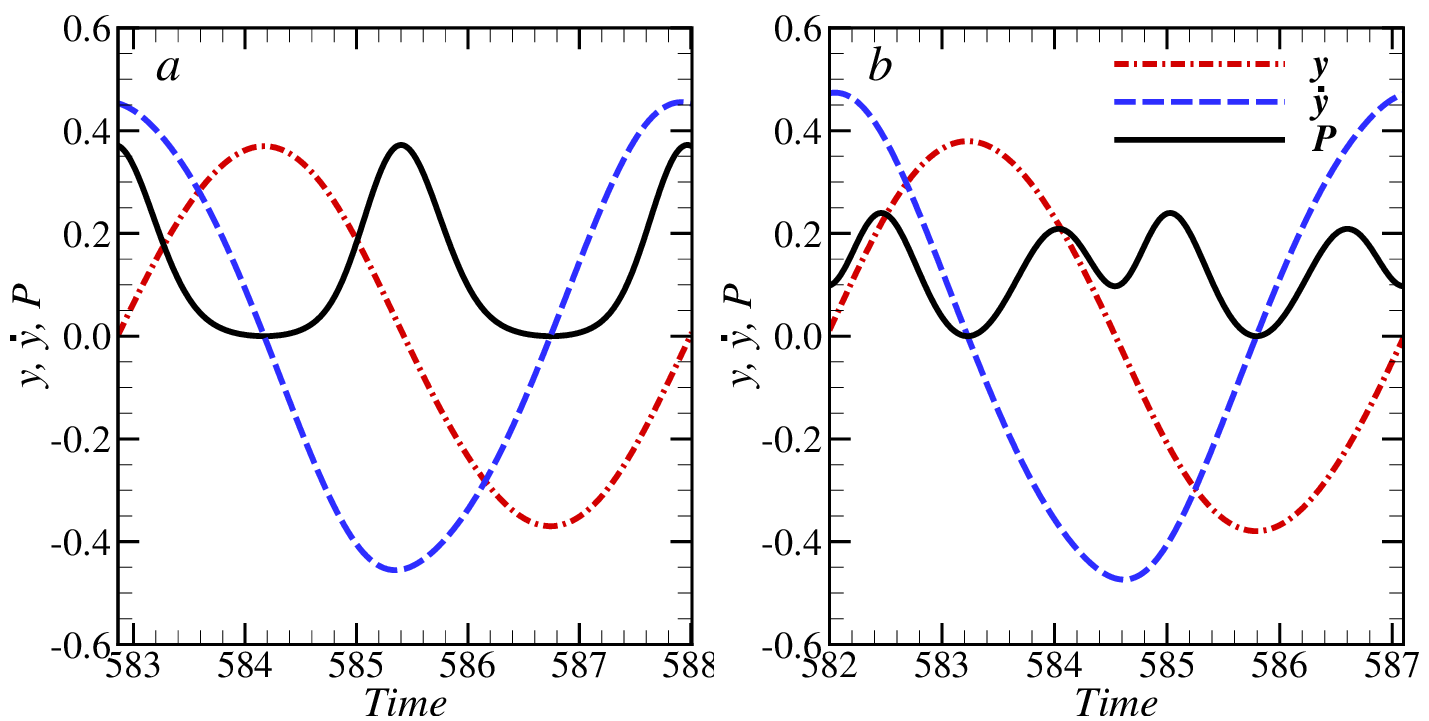}
\caption{\label{fig:pt_2c} Variation of power, position and velocity of the cylinder with time for (a) $y_c = 0.39$ at $\xi_{m0} = 6.8 \times 10^{-6}$ and (b) $y_c = 0.10$ at $\xi_{m0} = 1.0 \times 10^{-5}$ at $U_r$ = 5.2, $m$ = 2 and $Re = 150$ for two identical coils of $a$ = 0.6 and $L$ = 0.6.}
\end{figure}

The effect of the two $y_c$ values on power for mass ratio $ m = 2$ at lock-in ($U_r = 5.2$) is investigated. The Reynolds number is kept at 150, and the length and radius of the coils are taken as 0.6 and 0.6. The variations of power with time for the two cases are plotted in fig.~\ref{fig:pt_2c}. The plots corresponds to the their optimal damping situations. The instantaneous power for the $y_c = 0.39$ case is shown in fig.~\ref{fig:pt_2c}a. The peak value of power ($P_p$) is 0.39, and is generated when the cylinder is located near the centre. Although the frequency of power is the same as that of the CD case presented in section \ref{subsec:CvsEM}, i.e., $f_p$ = 2$f_c$, the temporal variation shows some dissimilarity. The power does not vary in a sinusoidal fashion in this case. In a cycle, the duration for which power is more than $P_p/2$ is smaller than the duration for which the power is less than $P_p/2$. The instantaneous power for the $y_c = 0.10$ case is shown fig.~\ref{fig:pt_2c}b. The peak power for this case is $P_p = 0.24$, which is smaller than that for the single coil EMD case presented in section \ref{subsec:CvsEM}. The peak power is generated when the cylinder displacement is $y = \pm 0.22$. Similar to the single  coil case, there are two unequal peaks for the power, and the lower and higher peaks occurs when the cylinder is moving towards and away from its mean position, respectively.  However unlike the single coil case, the power is not zero at the centre because $\xi_{m} \neq 0 $ there, and therefore $f_p$ = 2$f_c$ for this case.

\begin{figure}[h]
\centering 
\includegraphics [width= 0.7\textwidth] {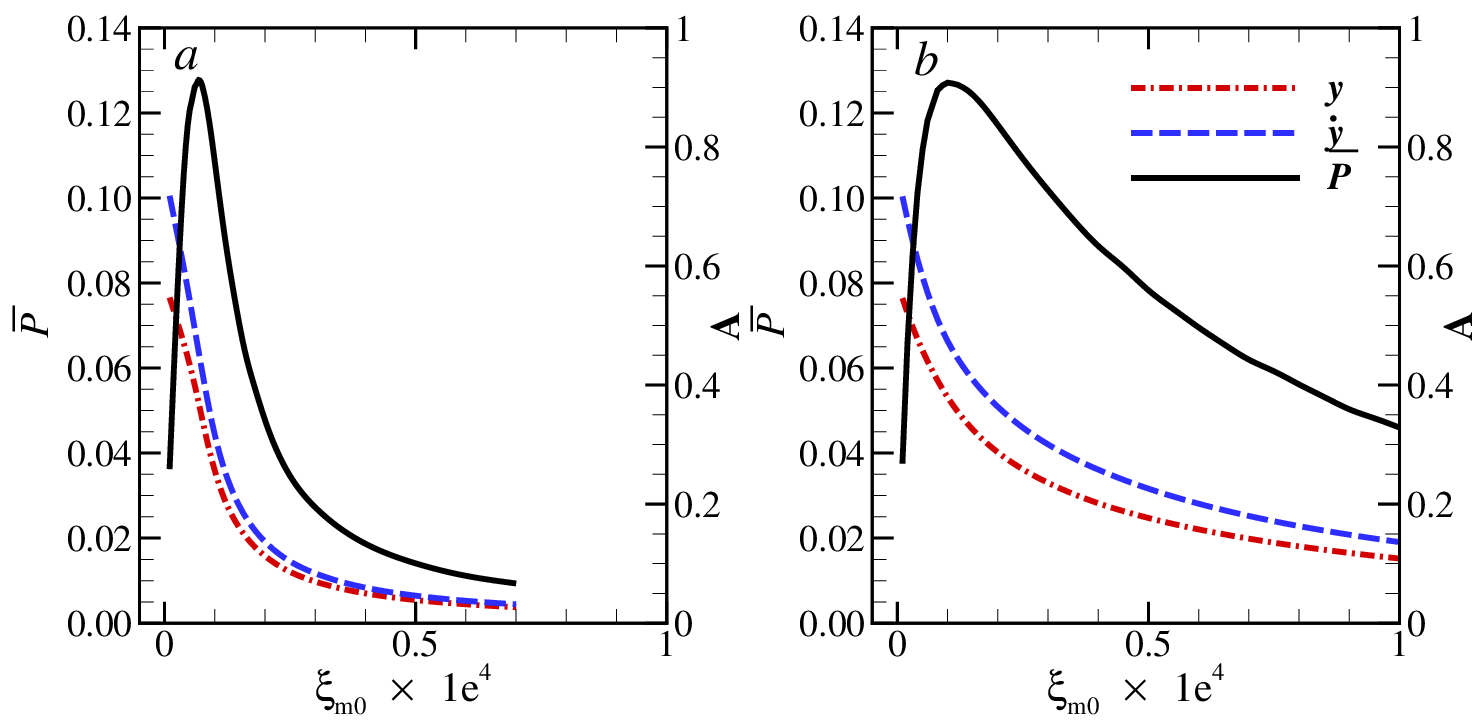}
\caption{\label{fig:p_vs_xi_2c} Variation of average power with $\xi_{m0}$ for (a) $y_c = 0.39$ and (b) $y_c = 0.10$ at $U_r$ = 5.2 and $m$ = 2 for two identical coils of $a$ = 0.6 and $L$ = 0.6.}
\end{figure}

In section \ref{subsec:CvsEM}, it was shown that the CD case had a lower $Q$ value compared to the single coil EMD case. Figure \ref{fig:p_vs_xi_2c}a shows the variation of average power with $\xi_{m0}$ for $y_c = 0.39$. The value of $Q$ for this case is 1.4, which is smaller than that for the CD case. The reason for this small value of $Q$ is the rapid decrease in the displacement and velocity amplitudes with $\xi_{m0}$, as seen in fig.~\ref{fig:p_vs_xi_2c}a. This occurs due to a higher value of $\xi_{m}$ at the centre, resulting in a strong damping force. The variation of average power for $y_c = 0.10$ with $\xi_{m0}$ is shown in fig.~\ref{fig:p_vs_xi_2c}b. The value of $Q$ in this case is 5.8, which is smaller than the one coil EMD case, and this can be attributed to the fact that $\xi_{m}$ is non-zero at the centre. Again, the maximum average power for both $y_c$ values is 0.13. The displacement amplitudes of the cylinder at optimal $\xi_{m0}$ for $y_c$ = 0.10 and 0.39 are 0.38 and 0.37, respectively, with corresponding cylinder velocity amplitudes of 0.47 and 0.46.

\section{Conclusions}
The problem of harvesting power from VIV of a circular cylinder was investigated numerically using a spectral-element based FSI solver. Specifically, the average power production was compared for constant damping (CD) ratio and a realistic electromagnetic induction-based magnet-coil energy transducer. The magnet-coil system was modelled as a damper having spatially varying electromagnetic damping (EMD) ratio. There was an optimal damping ratio, for both the systems, at which maximum power was harnessed. The results show that the temporal variation of instantaneous power is quite different for the two cases but, perhaps surprisingly, both systems produce the same maximum average power (non-dimensional value of 0.13 at $Re$ = 150). The EMD case produced a larger value of peak power as compared to the CD case. The frequency of the power signal for the EMD case was twice of that of CD case. In terms of the range of damping values for which a significant amount of power is extracted from the flow, the EMD setup is superior compared with the constant damping case.

The effects of coil length, radius and mass ratio were also explored and quantified. It was found that both length and radius of the coil do not affect the maximum average power although the average power versus damping ratio relation does change. The value of optimal damping ratio increases with increases in both length and radius of the coil. The increase in mass ratio increases the maximum average power by a small amount. The more significant effect of mass ratio is on the range of reduced velocity for synchronization of the cylinder. A smaller mass ratio cylinder is seen to have a larger synchronised region, and thereby can produce a significant amount of power over a large range of reduced velocities.

The effect of Reynolds number on power output was also studied over the range leading to two-dimensional periodic flow. The average power increases with an increase in Re. There was a 30\% and 45$\%$ increase in maximum average power when Re was increased from 100 to 150, and to 200, respectively. The increase in average power can be attributed to the combined increases in vibration amplitude and frequency of the cylinder with Re.

The possibility of using two coils positioned symmetrically about the centre of the transverse direction was also examined. Two values of the distance between the coils were studied. These distances were chosen such that the resultant damping ratios had different variations with spatial coordinate. Both the cases give different temporal variation of power due to the difference in the nature of damping. The case where damping was highest at the centre, produced the largest peak power. The average power decreased faster with the damping ratio for this case compared with the case where damping is not highest at the centre. The maximum average power was the same (non-dimensional value of 0.13 at $Re$ = 150) for both the cases. The systems with zero damping at the center were found to be less sensitive (have higher Q values) to variation in damping from it optimal value.

To summarise, the constant and electromagnetic damping cases produce entirely different temporal variations of instantaneous power output. In addition, the fundamental frequency of power is different between the two cases. On the other hand, the average power does have a strong dependence on the vibration amplitude of the cylinder, which depends on Reynolds number, despite showing little dependence on the nature of damping. Of potential importance is that the system quality, which measures the insensitivity of power output to choosing the optimal system damping, is considerably better with the spatial-varying damping associated with the electromagnetic damping system.

\section{Acknowledgements}
Support from Australian Research Council Discovery Grants DP0877327, DP110102141, DP110100434 and DP130100822, and computing time allocations from the National Computational Infrastructure (NCI), the Pawsey Supercomputing Centre with funding from the Australian Government and the Government of Western Australia, and the Monash MONARCH clusters are gratefully acknowledged.

\end{document}